\documentclass[11pt]{article}
\usepackage{moriond,epsfig}

\def\be{\begin{equation}}
\def\ee{\end{equation}}
\def\bea{\begin{eqnarray}}
\def\eea{\end{eqnarray}}

\begin{document}
\vspace*{4cm}
\title{MEASURING LEPTONIC CP VIOLATION WITH A WATER CERENKOV}

\author{E. COUCE}

\address{IFIC, Universidad de Valencia, \\
E-46100 Burjassot, Spain.}

\maketitle\abstracts{
In this talk, we present the physics case for a megaton Water Cerenkov
detector in addressing some of the still pending questions in neutrino oscillations physics,
particularly the measurement of leptonic CP violation. We compare different future beams 
that could profit from a water detector and analyse, for the case $\theta_{13} \approx 3^\circ$ (the 
limit that can be reached by under-construction experiments), the signal-to-background rate for
a $\beta$-beam setup with the radioactive ions accelerated to $\gamma = 350$.}

\section{Introduction}

Two parameters of the leptonic mixing matrix remain still essentially unknown: the mixing angle $\theta_{13}$, which we know
from reactor experiments lies below 11.5$^\circ$, and $\delta$, the CP-odd phase in the leptonic sector, of which we know nothing about. 
In the near future we can expect more stringent limits on the value of $\theta_{13}$, thanks to under-construction experiments\cite{Itow:2001ee}$^,\,$\cite{Ayres:2004js}, but these experiments can do little to answer the relevant question of the possible existence 
of CP violation in the leptonic sector. Thus
the search for leptonic CP-violation, together with the precise measurement of $\theta_{13}$ and the determination of the neutrino 
mass hierarchy (sign of $\Delta m^2_{23}$) are challenges that still need to be addressed by a future new generation of long-baseline,
high-precision neutrino oscillation experiments.

To address this challenge, these new experiments will need to measure very small oscillation probabilities at $E/L \approx \Delta m^2_{atm}$.
This can only be achieved with very large statistics and a tight control of the systematic errors, implying both the use of very 
intense and well-known $\nu$ beams and massive detectors with low systematic uncertainties.

\section{The experimental challenge}
Regarding the beams, there are three choices nowadays that appear to be the most promising candidates to measure these unknowns:
\begin{description}
\item[super-beam:] high-intensity conventional neutrino beams from pion decay (composed mostly of $\nu_\mu$, but also with contamination of 
$\bar\nu_\mu$, $\nu_e$ and $\bar\nu_e$);
\item[$\beta$-beam\cite{Zucchelli:sa}:] pure $\nu_e$ ($\bar\nu_e$) beam produced from accelerated $\beta$-radioactive ions;
\item[$\nu$-factory\cite{Geer:1997iz} beam:] $\nu_\mu$ and $\bar\nu_e$ ($\bar\nu_\mu$ and $\nu_e$) obtained from the decay of accelerated muons.
\end{description}
Notice that both $\beta$-beams and $\nu$-factory beams, unlike super-beams, are pure (without $\nu_\mu$ and $\nu_e$ contamination respectively),
and the energy dependence of their fluxes can be computed analytically. Also notice that for a $\nu$-factory experiment the determination 
of the charge of the detected lepton is mandatory.

The need for very high statistics favours the use of very massive detectors. One of the main limitation of the size of a detector comes from
the actual costs of the materials employed in their construction. Although water detectors might not be, intrinsically, the optimal choice for
the study of neutrino oscillations, they are undoubtedly the cheapest. A water-based detector could reach sizes at least one order of
magnitude above what any other technology might allow cost-wise.
In addition to this, Water Cerenkovs have already been studied extensively, and have played key roles in establishing much of what we
know today about neutrino oscillations. The 50-kton SuperKamiokande detector in Japan has been
conducting highly successful and relevant research in the field of neutrino physics since it became operational in April of 1996.
For all these reasons it is interesting to study how far this technology could take us in the future. In this paper we focus on the physics
case for a Water Cerenkov detector with a fiducial mass of 500 ktons (roughly 20 times the size of SuperKamiokande). 

Of the three beam types mentioned above, a Water Cerenkov could in principle be used only with super-beams and $\beta$-beams, since for 
a neutrino factory the charge of the detected $\mu$ has to be distinguished, and magnetising such a large detector is certainly 
a non-trivial matter\footnote{Nevertheless, it should be noted that some alternatives to magnetisation have been proposed,
such as the dissolution of Gd in a SuperKamiokande-like detector to make neutrons visible\cite{Beacom:2003nk}.}. 
Furthermore, in order to establish the limits of the Cerenkov detector technology, we will focus on the most promising of the two remaining beams, the 
$\beta$-beam. Unlike super-beams, $\beta$-beams do not suffer from flavour or sign contamination.

There have been many studies of $\beta$-beam experiments employing a Water Cerenkov detector 
(see for example \cite{Bouchez:2003fy}$^,\,$\cite{Burguet-Castell:2003vv}$^,\,$\cite{Burguet-Castell:2005pa}). 
It has been shown\cite{Burguet-Castell:2005pa} that 
the optimal setup for reaching the highest sensitivities to leptonic CP violation and $\theta_{13}$ consisted of a baseline of $\sim700$ km 
and the radioactive ions $^6$He and $^{18}$Ne accelerated to a $\gamma$ factor of 350. 
Therefore, this is going to be the setup considered in our present work.

\section{Signal and Background in a $\beta$-beam}
The $\beta$-beam is a beam of pure $\nu_e$ ($\bar\nu_e$) produced from $\beta$ radiation of accelerated radioactive ions. Nowadays the best choices 
for the radioactive ions appear to be $^{18}$Ne for the $\nu_e$ beam and $^6$He for the $\bar\nu_e$\footnote{The use of different sets of ions is
under study\cite{Donini:2006dx}.}. Out of all the possible measurements that could be
carried out with such a beam, the most sensitive one to $\theta_{13}$ and leptonic CP violation would be the so-called \emph{golden channel}: the 
study of the $\nu_e \rightarrow \nu_\mu$ ($\bar\nu_e \rightarrow \bar\nu_\mu$) oscillations.

There are two main analysis cuts that could be used to isolate the $\nu_\mu$ signal from the $\nu_e$ background:
\begin{description}
\item[PID cut:] In a water Cerenkov detector, the signal is characterised by the detection of a ring of light produced 
by Cerenkov effect, detected by the photomultipliers attached to the detector wall. The signal for a $\mu$ is in 
general much sharper and better defined than that of an $e$, so a study of the shape of the detected ring can provide 
valuable information of the ID of the detected particle. In this study we have considered a SuperKamiokande-like PID algorithm.
\item[Michel $e^-$ cut:] The $\mu$ often decay into $e$ that could themselves produce Cerenkov radiation detected by the 
photomultipliers. Thus our second analysis cut is the requirement of the observation of a delayed second ring, detected 
after a time interval related with the decay time of the $\mu$.
\end{description}

The energy reconstruction is done assuming quasi-elastic kinematics. However, for $\gamma = 350$ 
the quasi-elastic cross section is not dominant. Nevertheless the resulting reconstructed energy bias is not necessarily
the limiting factor of the analysis, as this effect can be studied and accounted for using Montecarlo simulations.

\begin{figure}[ht!]
\begin{center}
\epsfig{file=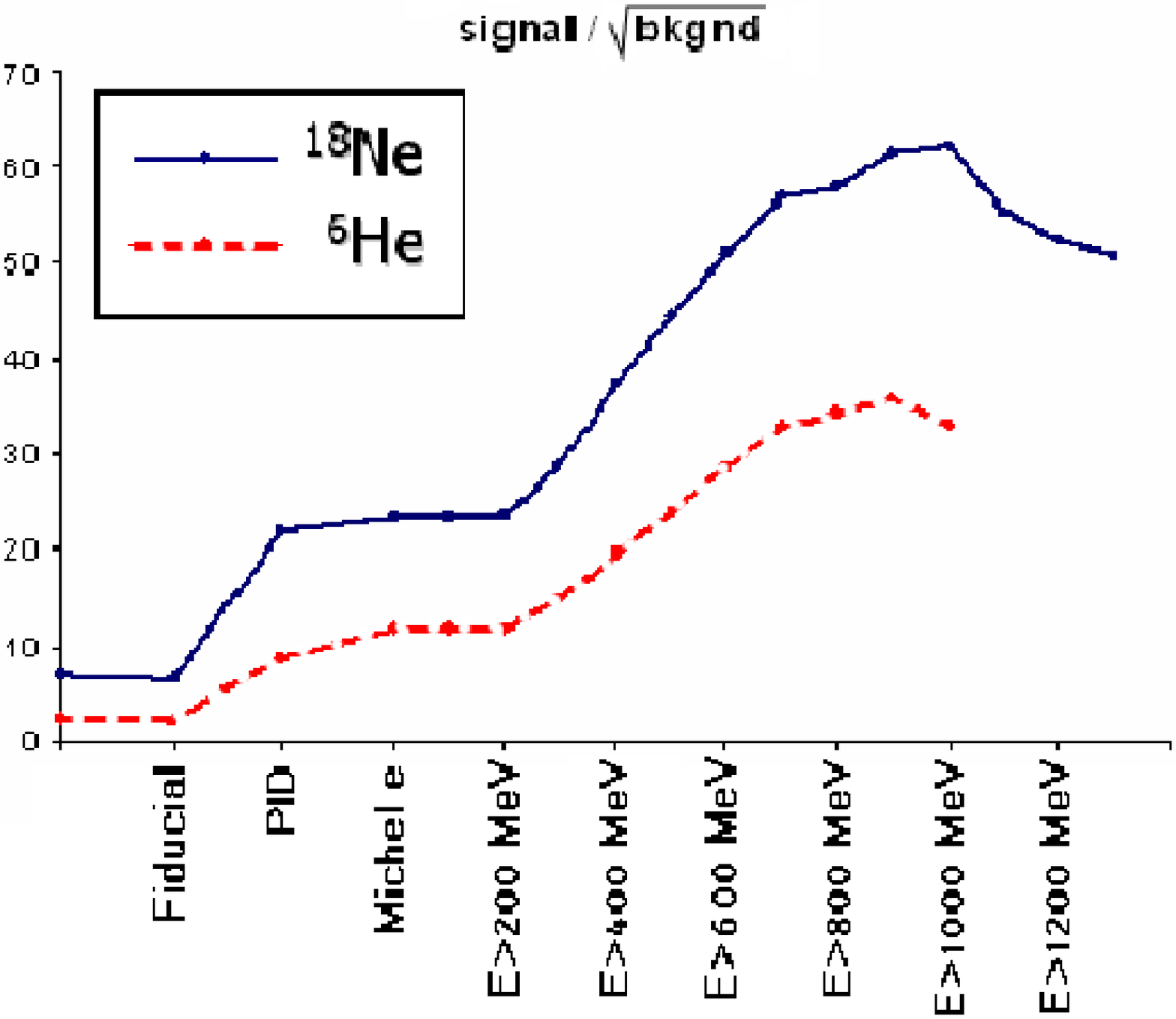,height=6.8cm}
\epsfig{file=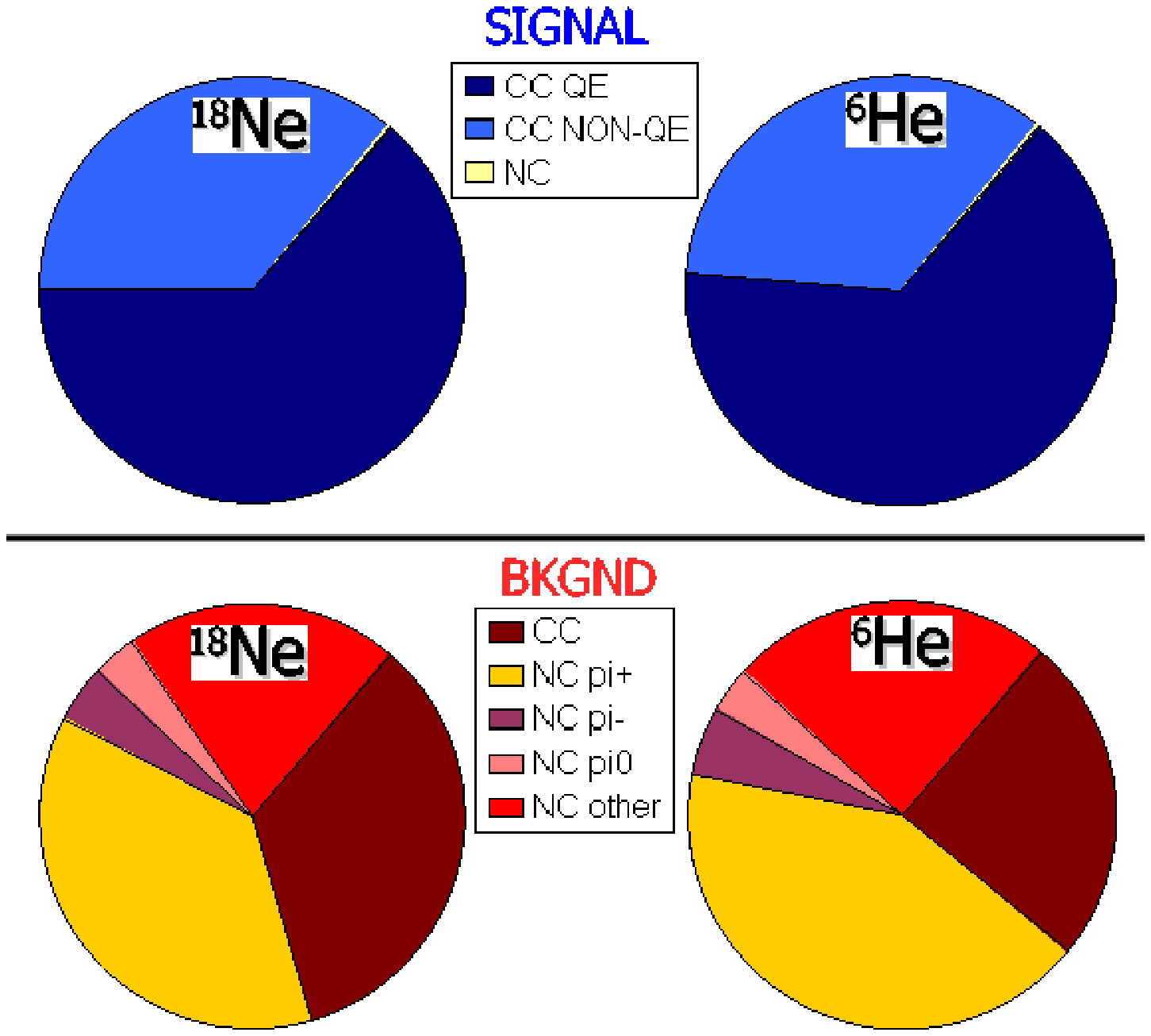,height=6.8cm}
\end{center}
\caption{\emph{Left:} Effect of each of the analysis cuts on the evolution of the signal-to-noise ratio for a $\gamma = 350$ $\beta$-beam. Also
shown are the effects of different minimum reconstructed-energy cuts. 
\emph{Right:} Final composition of signal ($\nu_\mu$ detected as $\nu_\mu$) and background ($\nu_e$ detected as $\nu_\mu$), after all analysis
cuts, for events with reconstructed energy above 500 $MeV$. For both plots a value of $\theta_{13} = 3^\circ$ has been assumed.}
\label{fig:comp}
\end{figure}

Left side of figure \ref{fig:comp} shows the effect of the analysis cuts on the signal-to-noise ratio 
when $\theta_{13} = 3^\circ$. As can be seen the ratio obtained after the Michel $e^-$ cut can be further 
increased by a factor 3 by the application of an additional cut on the minimal value of the reconstructed 
$\nu$ energy. Because of the assumption of quasi-elastic kinematics in energy reconstruction, a large part 
of the background (coming from NC events) is reconstructed at low energies and can be effectively eliminated 
with a minimum energy cut.

Right side of figure \ref{fig:comp} shows the signal and background final composition for events with a reconstructed 
energy above 500 MeV. As expected most of the signal is non-quasi elastic. The main contribution to the background 
is given by NC $\pi^+$ events and CC $\pi^+$, multi-$\pi$ and deep-inelastic events for which the electrons go undetected.

\section{The search for CP-violation}
\begin{figure}[ht!]
\begin{center}
\epsfig{file=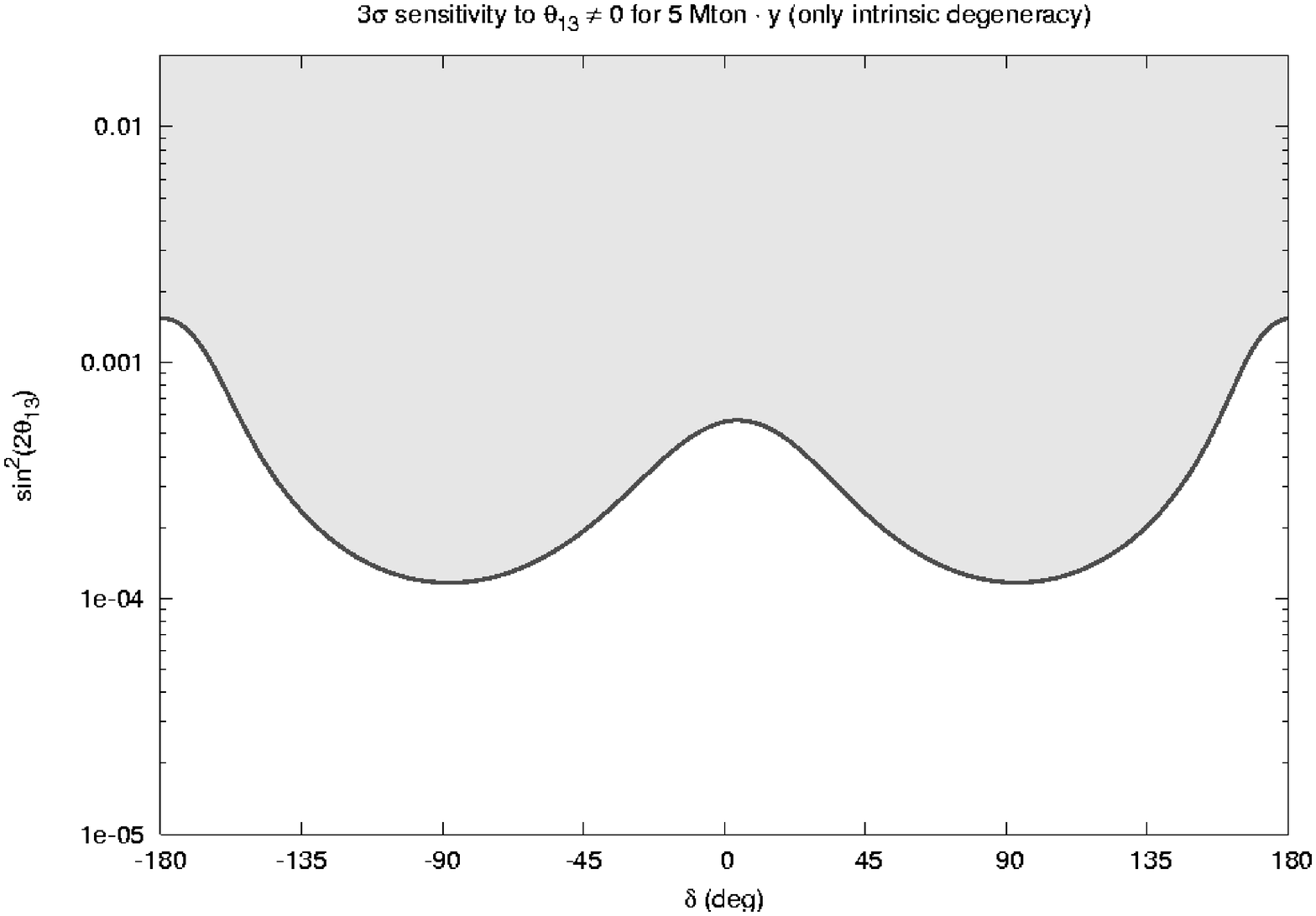,height=5.5cm}
\epsfig{file=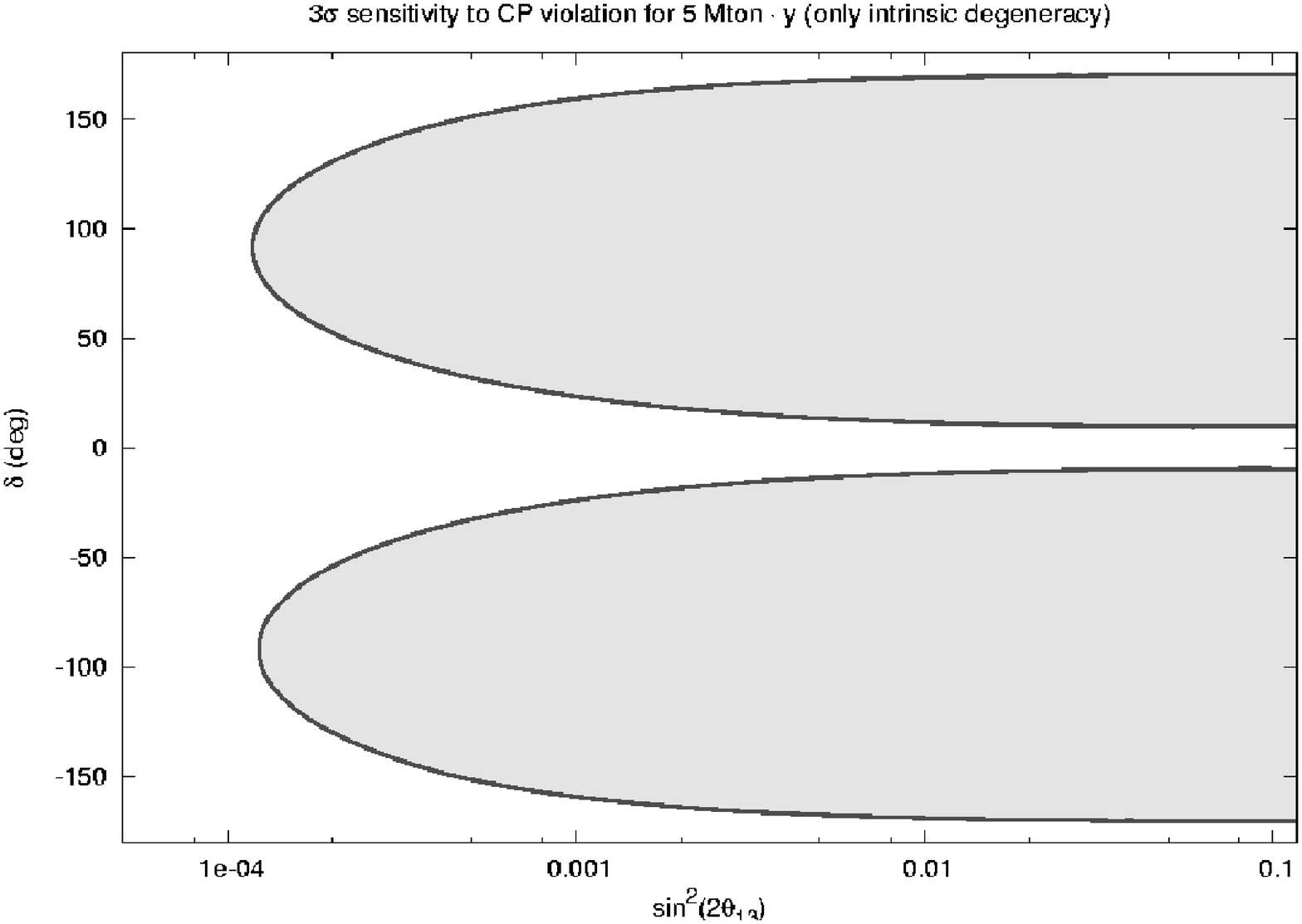,height=5.5cm}
\end{center}
\caption{3-$\sigma$ sensitivity to $\theta_{13}$ (left) and leptonic CP violation (right) of a beta-beam
setup with $\gamma = 350$ and $L=700$ km, for 5 Mton $\cdot$ year.}
\label{sensitivity}
\end{figure}

To conclude we present the preliminary results of our simulation for the measurement of $\theta_{13}$ and 
leptonic CP violation. Figure \ref{sensitivity} shows the exclusion plots for a 500-kton-fiducial-mass 
$\beta$-beam experiment, with $\gamma = 350$, a baseline of 700 km and a running time of 10 years. Intensities
of $2.9 \cdot 10^{18}$ for the $^6$He ions and of $1.1 \cdot 10^{18}$ for the $^{18}$Ne ions were considered.

If the values of $\theta_{13}$ and $\delta$ chosen by Nature lay in the shaded regions of the plots, the experiment 
would be able to establish a non-zero value for $\theta_{13}$ and leptonic CP violation respectively at 
3-$\sigma$ CL or better. For the calculations, we assume a global systematic error of 5\% on the signal, 10\% 
on the background, and 1\% on the neutrino-to-antineutrino cross section ratio.

The CP discovery potential of the $\beta$-beam setup presented here is competitive with that of a neutrino factory.

\end{document}